\documentclass[letter]{aa}  

\usepackage{graphicx}
\usepackage{graphbox}
\usepackage{xcolor}
\usepackage[colorlinks=true,citecolor=blue,linkcolor=blue,urlcolor=blue]{hyperref}

\usepackage{txfonts}
\newcommand{\Msun}{\,$M_{\odot}$}       
\newcommand{\Lsun}{\,$L_{\odot}$}       
\newcommand{\um}{\,$\mu$m}      
    
\newcommand{\CO}{CO(1--0)}

\begin{document}

   \title{The tight correlation of PAH and CO emission from $z\sim 0$ to 4
   }

   \subtitle{}

   \author{Irene Shivaei
          \inst{1}
          \and
          {Leindert A. Boogaard\inst{2,3}}
          }

   \institute{Centro de Astrobiolog\'{i}a (CAB), CSIC-INTA, Carretera de Ajalvir km 4, Torrej\'{o}n de Ardoz, 28850, Madrid, Spain\\
              \email{ishivaei@cab.inta-csic.es}
         \and
          Max Planck Institute for Astronomy, K\"{o}nigstuhl 17, 69117 Heidelberg, Germany
          \and
          Leiden Observatory, Leiden University, PO Box 9513, 2300 RA Leiden, The Netherlands \\
          \email{boogaard@mpia.de, boogaard@strw.leidenuniv.nl}
          }

   \date{Received XX; accepted XX}

 
  \abstract
   {}
   {The cold molecular gas mass is one of the crucial, yet challenging parameters in galaxy evolution studies. Here, we introduce a new {calibration and} method for estimating molecular gas masses using the mid-infrared (MIR) {photometry}. This topic is timely, as the \emph{James Webb Space Telescope} (\emph{JWST}) now allows us to detect the MIR emission of typical main-sequence galaxies across a wide range of masses and star formation rates with modest time investments. Additionally, this Letter highlights the strong synergy between ALMA and {\em JWST} for studies of dust and gas at cosmic noon.}
   {We combine a sample of 14 main sequence galaxies at $z=1-3$ with robust CO detections and multi-band MIR  photometry, along with a literature sample at $z=0-4$ with CO and PAH spectroscopy, to study the relationship between PAH, CO(1-0), and total IR luminosities. PAH luminosities are derived from modeling a wealth of rest-frame UV to sub-mm data.  The new $z=1-3$ sample extends previous high-$z$ studies to {about} an order-of-magnitude lower PAH and CO luminosities, into the regime of local starbursts for the first time.}
   {The PAH-to-CO luminosity ratio remains constant across a wide range of luminosities, for various galaxy types, and throughout the explored redshift range. In contrast, the PAH-to-IR and CO-to-IR luminosity ratios deviate from a constant value at high IR luminosities. The intrinsic scatter in the L(PAH)-L$^{\prime}$(CO) relation is 0.21\,dex, with a median of 1.40, and a power-law slope of $1.07\pm 0.04$. Both the PAH-IR and CO-IR relations are sub-linear. Given the tight and uniform PAH-CO relation over $\sim 3$ orders of magnitude, we provide a recipe to estimate the cold molecular gas mass of galaxies from PAH luminosities, with a PAH-to-molecular gas conversion factor of $\alpha_{\rm PAH_{7.7}} = (3.08 \pm 1.08) (4.3/\alpha_{\rm CO})$\,\Msun/\Lsun. This method opens a new window to explore the gas content of galaxies beyond the local Universe using multi-wavelength {\em JWST}/MIRI imaging.}
   {}


   \maketitle
%

\begin{figure*}[ht]
        \centering
        \includegraphics[width=.8\textwidth]{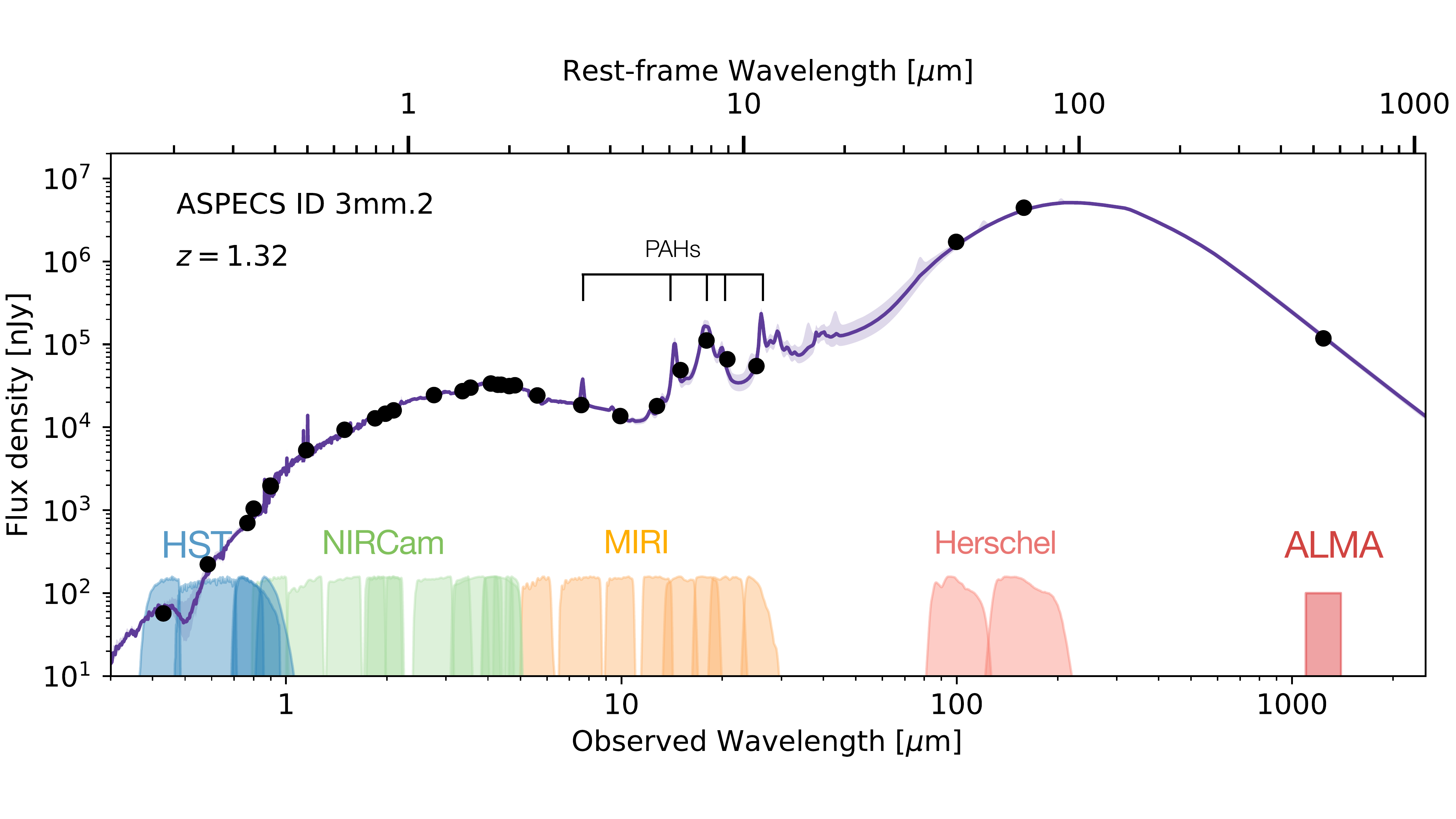} 
    \caption{Demonstration of the data from \emph{HST}, \emph{JWST}, \emph{Herschel} and ALMA, and the best-fit Spectral Energy Distribution (SED), for one of the galaxies in the sample. The wealth of photometric observations from rest-frame UV to FIR, in addition to spectroscopic redshift, enables detailed SED fitting that can robustly constrain the PAH features. The filter transmission curves of the available photometric data for the sample are shown at the bottom. The SEDs of the full sample will be available online through the journal.}
    \label{fig:sed}
\end{figure*}

\section{Introduction}

Stars form from clouds of molecular hydrogen in the cold and dense regions of the interstellar medium (ISM; \citealt{mckee07}), and their radiation interacts with the gas and dust in the ISM through photoexcitation, photoionization, photodissociation, and heating processes. These processes determine the physical state of the ISM and determine the shape of the gas and dust spectral energy distribution (SED) of galaxies. Therefore, simultaneous study of these components and their relationship is an integral part of understanding galaxy evolution.
To study the molecular gas content of galaxies, CO emission is often adopted as a tracer. This is simply because the emission from cold H$_2$, the main constituent of the molecular gas in galaxies, is weak and extremely challenging to observe \citep{bolatto13}.
Different rotational transitions of CO emit at millimeter (mm) wavelengths, which are observable with (sub)-mm facilities, such as ALMA, over a wide range of redshifts \citep{carilli13}. 
On the other hand, to study dust, observing its emission in infrared (IR) is required. In the mid-IR (MIR) regime, the emission emerges from small dust grains that, in star-forming galaxies, are dominated by polycyclic aromatic hydrocarbons (PAHs) in the photodissociation regions. PAHs are complex organic molecules that absorb UV photons and radiate in MIR, with the strongest emission at 7.7{\um} \citep{tielens08}. As the UV radiation in the local radiation field of star-forming galaxies is often dominated by radiation from young stars, PAH emission is often used as a star-formation rate (SFR) indicator {\citep{calzetti11,rujopakarn13,shivaei17,whitaker17}}. This SFR diagnostic has the advantage of being insensitive to dust attenuation, compared to UV and optical diagnostics \citep{kennicutt12}, but also the caveat of having a non-linear relation at low metallicities \citep{shivaei17} and high IR luminosities \citep{shipley16}. On the other hand, PAHs can also get excited by nonionizing radiation from evolved stellar populations \citep{draine01,sellgren01,haas02,li02,zhang22}, which also heats the diffuse interstellar dust and gas.  In fact, 
PAHs are also observed to be co-spatial with cold dust and cold gas in local galaxies \citep{chown21,gao22,leroy23a,zhang23}. 

While many studies use the CO and PAH observations to study molecular gas mass and SFR, respectively, fewer have directly studied the relationship between the PAH and CO emission. Among these few studies, it has been shown that PAH emission correlates strongly with CO emission at sub-kpc scales over diverse environments of star-forming and AGN host local galaxies \citep{chown21,gao22,leroy23b,zhang23} and that the PAH-CO correlation is stronger than that of CO-SFR or PAH-SFR \citep{whitcomb23}. Beyond the local Universe, there are even fewer studies that have looked into the PAH-CO connection \citep{pope13, cortzen19}. \cite{cortzen19} compiled a large sample of galaxies with MIR spectroscopy and CO observations and find that on integrated scales, there is a universal PAH-CO relation from $z\sim 0$ to $\sim 4$. The higher redshift datasets used in \cite{cortzen19} are inevitably limited to extreme galaxy populations of sub-mm galaxies (SMGs), ultra-luminous IR galaxies (ULIRGs), starbursts, or very massive/star-forming main-sequence galaxies,
as the only sufficiently sensitive MIR spectrometer before {\em JWST} was {\em Spitzer} IRS. As a result, their results in the LIRG regime and below are based on $z\sim 0$ data alone, missing the typical, $L_*$ population of galaxies at high redshifts. This limitation can now be overcome by the sensitive Mid Infrared Instrument (MIRI; \citealt{rieke15,wright23}) onboard the {\em JWST}.

The advent of {\em JWST} enables us to probe PAH emission in typical galaxies at cosmic noon though both spectroscopy and photometry with MIRI. While spectroscopy is the ideal method, it is still very expensive even with MIRI medium-resolution spectrometer (MRS; for example, the JWST Cycle 3 program PAHSPECS, PID 5279, takes $\sim 10$ hours per object for typical main-sequence galaxies at $z\sim 1$). Thankfully, owing to the continuous multi-band coverage of the MIRI Imager, it is possible to robustly determine the luminosity of the broad PAH features at $z<3$, when observed in multiple bands of MIRI at 5.6 to 25.5\,{\um}. This has been demonstrated in \cite{shivaei24} using the largest available MIRI multi-band survey, SMILES \citep{rieke24,alberts24}.

In this Letter, we take advantage of the PAH measurements from the {\em JWST}/MIRI SMILES survey from \cite{shivaei24} and the CO measurements from the ALMA Spectroscopic Survey in the HUDF (ASPECS) from \cite{boogaard19,boogaard20}, to investigate the coupling between PAH and molecular gas emission in typical galaxies at $z=1-3$. To place our results within the broader context of the PAH and CO redshift evolution, we incorporate $z\sim 0$ data from the literature. Additionally, to explore the relationship across different galaxy populations, we include existing samples of starbursts, ULIRGs, and SMGs at $z=1-4$ from the literature. Thus, this Letter has two primary goals: 1) to extend the PAH-CO relationship to typical galaxy populations at high redshifts for the first time, and 2) to demonstrate the effectiveness of MIRI multi-band photometry in accurately measuring PAH luminosities, consistent with spectroscopic measurements. This is a particularly timely topic, as it opens up a new avenue for measuring molecular gas masses in statistically large and representative samples of galaxies using MIR imaging with the {\em JWST}.

\begin{figure*}[ht]
        \centering
        \includegraphics[align=c,width=.5\textwidth]{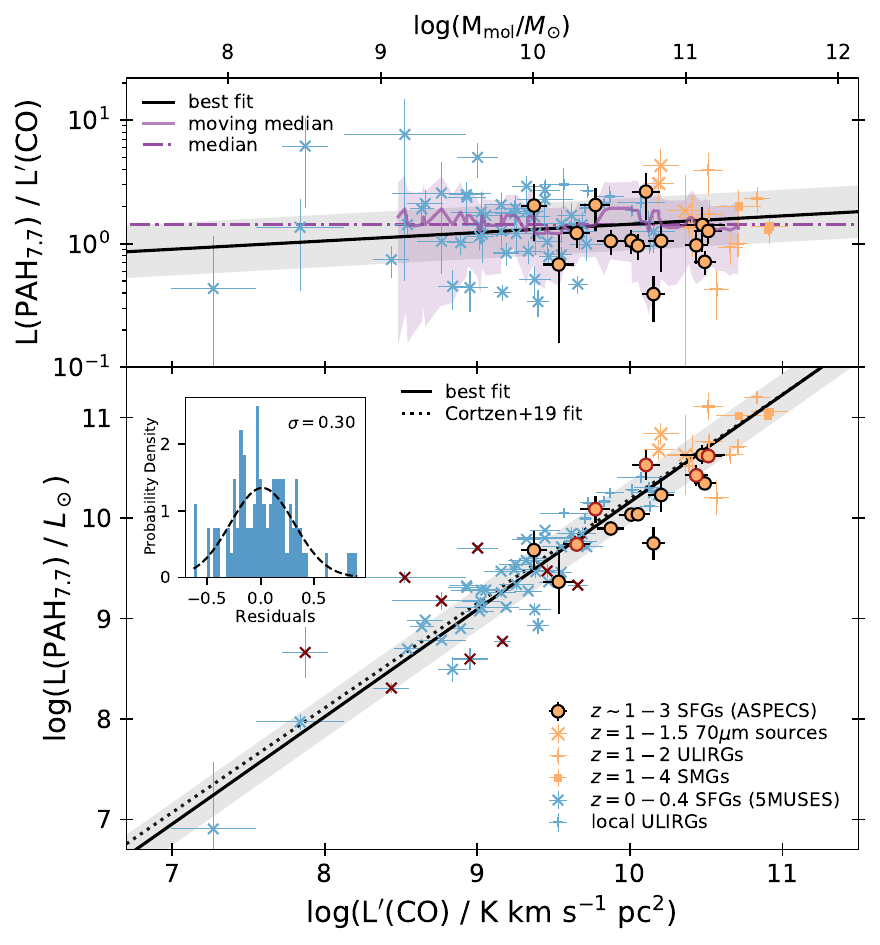} 
    \caption{\label{fig:co-pah} PAH and CO luminosity from $z\sim 0$ to 4. Data at redshifts below and above 1 are shown in blue and orange, respectively. The ASPECS sample from this work is shown with orange circles. Other samples are from the compilation of \citet{cortzen19} {: high-$z$ non-ULIRG sources \citep{pope13}, high-$z$ ULIRGs \citep{yan10}, local SFGs \citep{kirkpatrick14,cortzen19}, local ULIRGs \citep{armus07,desai07}.} Our sample clearly bridges the previously existing gap between the low and high-$z$ relations and extends the high-$z$ sample to {about} an order of magnitude lower PAH and CO luminosities, into the regime of local (U)LIRGS.
    We marked the AGN-dominated star-forming galaxies (SFGs) at all redshifts in red. For reference, molecular gas mass estimates from CO luminosities (assuming $\alpha_{\rm{CO}} =4.3$\,\Msun(${\rm K~km~s}^{-1}~{\rm pc}^{2})^{-1}$; \citealt{bolatto13}) are shown the top horizontal axis.
    In both panels, the best-fit linear model to log(L(PAH))$-$log(L$^{\prime}$(CO)) using \textsc{HyperFit} method \citep{robotham2015} is shown with a solid black line. Grey shaded region indicates the intrinsic vertical scatter from the fit. The inset panel shows the histogram of the fit residuals, which are centered at 0 with a small scatter ($\sigma$). 
    The C19 best-fit line is shown with a dotted line in the bottom panel. 
    The top panel also shows the median of L(PAH) to L$^{\prime}$(CO) ratio (dashed-dotted line) and its moving (running) median and scatter (purple solid line and shaded region). All statistical parameters are listed in Table~\ref{tab:fit}.
    As shown by the {moving} median and the fit, the PAH-to-CO luminosity ratio is fully consistent with a constant value for a diverse population of galaxies from SFGs to ULIRGs, SMGs, and starbursts at $z=0$ to $\sim 4$.}
\end{figure*}

\begin{figure*}[ht]
        \centering
        \includegraphics[width=.45\textwidth]{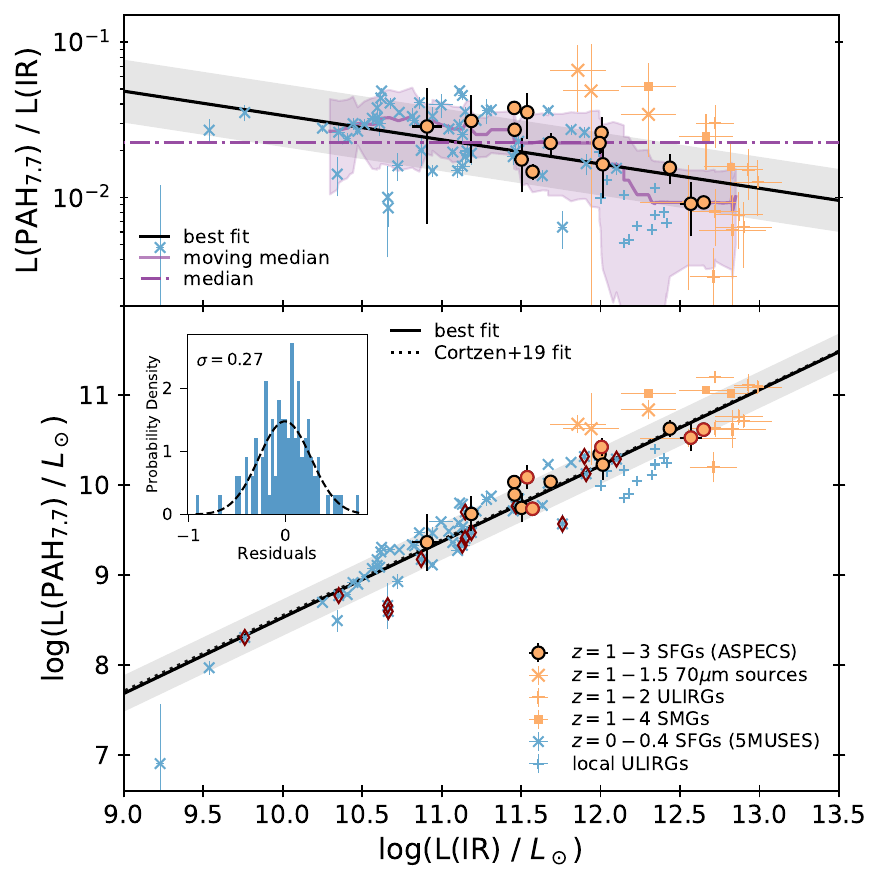} \quad
        \includegraphics[width=.45\textwidth]{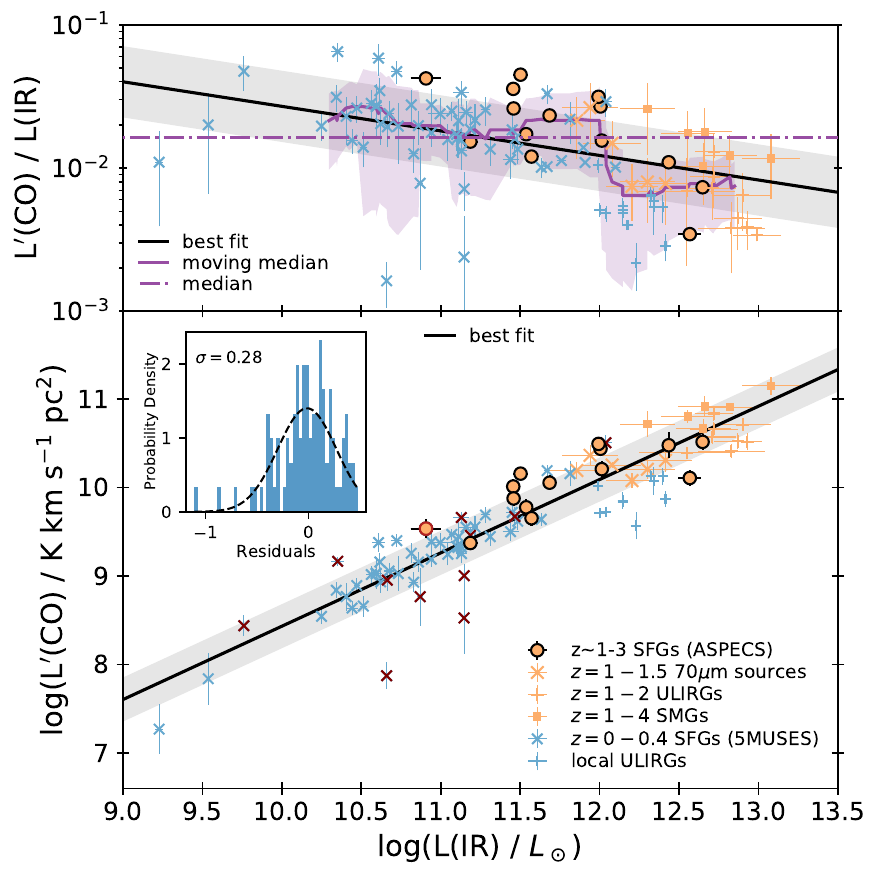} 
    \caption{Same as Figure~\ref{fig:co-pah} for the PAH-IR luminosity (left) and CO-IR luminosity relationships (right). Total IR luminosity (L(IR)) is calculated from 8 to 1000{\um} from the best-fit SEDs (Figure~\ref{fig:sed}).
    Unlike the CO-PAH relation in Figure~\ref{fig:co-pah}, the PAH-IR and CO-IR luminosity ratios are not constant across the population---they decrease at high IR luminosities (ULIRG regime), leading to sub-linear fits.}
    \label{fig:lir}
\end{figure*}

\section{Sample and Measurements}\label{sec:method}

\subsection{This work}
This paper is based on the the extensive observations in the Hubble Ultra Deep Field from the ALMA ASPECS and {\em JWST} SMILES surveys. The ASPECS large program performed spectral scans of the full ALMA bands 3 and 6 to provide an inventory of the cosmic molecular gas and dust content of galaxies out to high-redshift \citep{decarli19,decarli20}. SMILES is a 30-arcmin$^2$ MIRI multi-band imaging survey from JWST Cycle 1 GTO time \citep{rieke24,alberts24}. It observed in eight MIRI filters from 5.6 to 25.5{\um}, as shown in Figure~\ref{fig:sed}: F560W, F770W, F1000W, F1280W, F1500W, F1800W, F2100W, F2550W.

The sample in this work consists of {all but one (14 out of 15) of the} CO emitters in the flux-limited sample from ASPECS at $z=1$--3 \citep{boogaard19}.
 source (3mm.08) was excluded because it is almost fully behind a foreground spiral galaxy prohibiting an analysis of its SED. The redshift criterion was set to ensure sufficient coverage of the PAH MIR emission by MIRI and only one galaxy (3mm.13 {at $z=3.6$}) {from the full parent CO sample (16 sources)} was excluded by this criterion. All sources have extensive {\em HST} ACS and WFC3 (CANDELS; \citealt{grogin11, koekemoer11}) and {\em JWST} NIRCam (JADES; \citealt{rieke23}) coverage, and unambiguous far-IR {\em Herschel} counterparts \citep{elbaz11}.

For details about the {\em HST} and {\em JWST} photometry and SED fitting, we refer to \citet{shivaei24}.
The only difference between the SED fitting in \citet{shivaei24} and this work is the inclusion of FIR continuum data from {\em Herschel} and ALMA in this work. The FIR photometry is detailed in \citet{boogaard20} and includes the PACS photometry at 100 and 160\,\um\ (as the SPIRE photometry is typically strongly blended) as well as ALMA continuum at 1.2 and 3.0\,mm.  In brief, the SED fitting was done using the PROSPECTOR code \citet{johnson21}, assuming delayed-tau star formation history, a flexible Calzetti attenuation curve, \citet{DL07} IR models, and including nebular and AGN emission. All galaxies have spectroscopic redshifts from CO observations. An example of the best-fit SED model is shown in Figure~\ref{fig:sed}.

AGN are identified through X-ray emission from deep Chandra observations \citep{luo17,boogaard19} and SED fitting.  From the SED fitting, we identified two galaxies where the AGN contributed significantly to the IR SED (AGN fraction $> 0.05$; 3mm.01 and 3mm.09). All AGN are marked in the following figures, however, they do not significantly differ from the rest of the sample and excluding them do not change the results.

The PAH emission adopted in this work was measured from the best-fit SEDs. \cite{shivaei24} shows that a full MIR photometric coverage with MIRI bands provide reliable measurements for the broad PAH bands at cosmic noon. We limit our analysis to the 7.7{\um} PAH feature, as it is the strongest PAH feature and best constrained by photometry.
We measured the PAH 7.7{\um} luminosity for individual fits using Equation 19 of \citet{draine21}. In brief, we first integrated the flux density from 6.9 to 9.7{\um}, assuming the feature strength to be zero on either side. {This method of estimating the continuum is closest to the PAH decomposition method, such as those employed by codes like PAHFIT \citep{smith07} that are adopted in our literature comparison sample (Section~\ref{sec:literature}). We note, however, it can differ by a factor of 3.5 from estimates derived using cubic spline methods, which are known to underestimate PAH fluxes \citep{smith07, pope08b}. As the derived values are for the continuum-subtracted $7.7+8.6${\um} PAH complex luminosity, we then applied a 15\% correction for the 8.6{\um} feature contamination, to estimate the 7.7{\um} luminosity alone. This correction factor is determined from the median 7.7-to-8.6{\um} luminosity ratio in \cite{smith07}. Finally, we did not correct the PAH spectra for silicate absorption, as the silicate absorption bands tend to be weak or absent in typical star-forming regions \citep{brandl06,garcia-bernete22,donnan24}. Supporting this, our fits using the \cite{DL07} PAH models show no significant residuals around the 9.7{\um} silicate absorption. Future studies with both photometric and spectroscopic MIR data for normal star-forming galaxies at high redshifts will be needed to thoroughly compare the photometrically and spectroscopically derived PAH luminosities.}

For details about the CO measurements and conversion of the higher-$J$ CO observations to CO(1--0) we refer to \citet{boogaard19,boogaard20}, respectively.  In brief, we used low-$J$ CO(2--1) and CO(3--2) measurements from ALMA \citep{boogaard19,gonzalez-lopez19} or, where available at high-S/N, direct CO(1--0) measurements from the VLA \citep[that is, for 3mm.1,\,7 and 9]{boogaard20,riechers20}.  To convert the observed CO(2--1) and CO(3--2) transitions to {\CO}, we adopted the average conversion factors derived for these galaxies by \cite{boogaard20} of $r_{21}=0.75\pm0.11$ and $r_{31}=0.80\pm0.08$, respectively. 
The redshifts, luminosities and AGN classification are provided in Table~\ref{tab:data}.

\subsection{Literature} \label{sec:literature}
The literature comparison used in this work is adopted from the pre-{\em JWST} compilation of integrated PAH and CO measurements in \citet[][hereafter C19]{cortzen19}. In the absence of multi-band MIR photometry (before {\em JWST}), the only robust way of measuring PAH emission was spectroscopy. C19 incorporates samples with both CO and PAH spectroscopy from $z\sim 0$ to 4. However, owing to the sensitivity limitations, the high-redshift sample ($z>1$) is inevitably limited to SMGs, ULIRGs, starbursts, and massive galaxies. Here, we briefly describe the sample and refer to C19 for more detail.
At $z<1$, two samples of 5MUSES and local ULIRGs were adopted. 5MUSES \citep{cortzen19,kirkpatrick14} is a 24{\um} limited sample at $z=0.03-0.36$ with CO(1-0) measurements and {\em Spitzer} IRAC, MIPS, IRS, and far-IR {\em Herschel} SPIRE photometry. The AGN in this sample are identified from 6.2{\um} equivalent widths. The local ULIRG sample has IRS spectra from \citet{armus07} and \citet{desai07}. 
At $z>1$, there are nine 24{\um} selected ULIRGs at $z\sim 1-2$ from \citet{yan10} and twelve galaxies (SMG, BzK, 70{\um} selected) from various surveys compiled in \cite{pope13}. From that sample, we removed the three (out of six) MIPS 70{\um}-detected sources that are reported to be close pairs in optical but unresolved in IR and CO observations \citep{pope13} due to the uncertainty of MIR, FIR, and CO emission being co-spatial (i.e., from the same source). For the {higher-redshift sources} with only higher-$J$ lines, CO(1--0) luminosities are estimated using the conversion factors from \citet[][$r_{21} = 0.84\pm0.13$, $r_{31}=0.52\pm0.09$, $r_{41}=0.41\pm0.07$]{bothwell13}, {derived from $z=1$--4} SMGs \citep[cf.][]{birkin21}.

\begin{table*}
\caption{Statistical properties and best-fit relations between the PAH, CO and IR luminosities.} \label{tab:fit}           
\begin{tabular}{c c c c c c c c} 
\hline\hline  
{$y-x$} & {sample ($N$)} & {$\rho$} & {$p$-value} & {$\alpha$} & {$\beta$} & {$\sigma_{\rm int}$ [dex]} & {median}\\
\hline \\
$L({\rm PAH_{7.7}})/L_{\odot}-L^{\prime}{(\rm CO)/{\rm K~km~s^{-1}~pc^2}}$ & {All (86)} & {0.94} & {6e-40} & {$1.07\pm 0.04$} & {$-0.52\pm0.36$} & {0.21$\pm 0.02$} & {1.40$\pm 0.49$} \\
$L({\rm PAH_{7.7}})/L_{\odot}-L^{\prime}{(\rm CO)/{\rm K~km~s^{-1}~pc^2}}$ & {no AGN (71)} & {0.95} & {1e-37} & {$1.07\pm 0.04$} & {$-0.40\pm0.38$} & {0.19$\pm 0.03$} & {1.42$\pm 0.45$} \\
$L({\rm PAH_{7.7}})/L_{\odot} - L^{\prime}{(\rm CO)/{\rm K~km~s^{-1}~pc^2}}$ & {$z>1$ (28)} & {0.85} & {8e-9}& {$1.18\pm 0.18$} & {$-1.76\pm1.62$} & {0.19$\pm 0.07$} & {1.32$\pm 0.40$}\\
\hline \\
$L({\rm PAH_{7.7}})/L_{\odot}-L({\rm IR})/L_{\odot}$ & {All (95)} & {0.94} & {2e-45} & {$0.84\pm 0.02$} & {$0.09\pm0.23$} & {0.20$\pm 0.02$} & {$0.022\pm 0.010$}\\
$L({\rm PAH_{7.7}})/L_{\odot}-L({\rm IR})/L_{\odot}$ & {no AGN (76)} & {0.94} & {5e-36} & {$0.83\pm 0.02$} & {$0.31\pm0.28$} & {0.21$\pm 0.02$} & {$0.024\pm 0.010$}\\
$L({\rm PAH_{7.7}})/L_{\odot}-L({\rm IR})/L_{\odot}$ & {$z>1$ (29)} & {0.85} & {4e-9} & {$0.81\pm0.07$} & {$0.58\pm0.80$} & {0.19$\pm 0.04$} & {$0.018\pm 0.009$}\\
\hline \\
$L^{\prime}{(\rm CO)/{\rm K~km~s^{-1}~pc^2}}-L({\rm IR})/L_{\odot}$ & {All (92)} & {0.92} & {1e-38} & {$0.82\pm 0.02$} & {$0.15\pm0.29$} & {0.25$\pm 0.02$} & {$0.017\pm 0.008$} \\
$L^{\prime}{(\rm CO)/{\rm K~km~s^{-1}~pc^2}}-L({\rm IR})/L_{\odot}$ & {no AGN (76)} & {0.94} & {1e-36} & {$0.80\pm 0.03$} & {$0.43\pm0.32$} & {0.22$\pm 0.02$} & {$0.017\pm 0.007$} \\
$L^{\prime}{(\rm CO)/{\rm K~km~s^{-1}~pc^2}}-L({\rm IR})/L_{\odot}$ & {$z>1$ (31)} & {0.86} & {5e-10} & {$0.67\pm0.06$} & {$2.15\pm0.78$} & {0.19$\pm 0.04$} & {$0.013\pm 0.006$} \\

\hline 
\end{tabular}
Columns are, from left to right: pair of parameters {($y, x$)}, sample for which the fits are done and its size ($N$), Pearson correlation factor ($\rho$) and associated $p$-value for the $\log(y)-\log(x)$, slope {($\alpha$)}, intercept {($\beta$)}, and vertical intrinsic scatter from linear fit to $\log(y) = \alpha \log(x)+\beta$ (using HyperFit, errors are estimated using MCMC method; $\sigma_{\rm int}$), median and MAD of $y/x$. 
\end{table*}

\section{Results: PAH, CO and IR luminosity relationships}\label{sec:results}

In this section we explore the relationship between PAH, CO, and total IR luminosity using various statistical methods, including the Pearson correlation coefficient, linear regression, and calculating medians and scatters.  
To quantify the relationship, we fit the data using \textsc{HyperFit}\footnote{\url{https://github.com/CullanHowlett/HyperFit}}, which implements a two-dimensional model that takes into account intrinsic scatter in the relationship as well as heteroscedastic errors on both variables \citep{robotham2015}.

Figure~\ref{fig:co-pah} shows the relationship between the PAH 7.7{\um} emission and {\CO} luminosity from $z\sim 0$ to 4. The newly added data from this work extends the PAH and CO luminosity of previous studies at $z>1$ to {about} an order of magnitude lower values, into the regime of nearby galaxies, clearly bridging the gap between the $z\sim 0$ and the $z>1$ samples before {\em JWST}.
This extension reveals that the relationship between PAH and CO luminosity holds over the entire explored redshift and galaxy population ranges. As shown in the top panel of Figure~\ref{fig:co-pah}, the ratio of PAH-to-{\CO} luminosity stays constant over the full sample without any significant deviation from the median value of $\frac{\rm L(PAH_{7.7})}{\rm L^{'}(CO)} = 1.40\pm 0.49$, where 0.49 is the mean absolute deviation (MAD) of the ratio.  From the best-fit relation, the intrinsic vertical scatter is 0.21\,dex, while the scatter normal to the plane is 0.15\,dex, which are relatively small compared to most galaxy scaling relations {(e.g., $\sim 0.3$\,dex in the star formation law as in \citealt{kennicutt21}, and $\sim 0.3$\,dex in the star forming main-sequence relation at $z\sim 2$ as in \citealt{shivaei15})}. 
Our fit to the full sample agrees well with the linear relation of C19.  The best-fit slope being slightly super-linear is primarily driven by the 70\,$\mu$m-selected sources that all have high PAH-to-CO ratios with small error bars.  These sources are known to have very strong PAH emission {\citep{pope13,cortzen19}}, with the PAHs contributing a significant fraction of the total IR luminosity, as seen in their PAH-to-LIR ratio (Figure~\ref{fig:lir}).  Even so, the {moving} median of the PAH-to-CO ratio remains flat over the full luminosity range. The slope, intercept, and intrinsic scatter of the fit are given in Table~\ref{tab:fit}.

Figure~\ref{fig:lir} shows the PAH and CO luminosity relations with total IR luminosity, L(IR). While the relations are linear with small scatter, similar to that of PAH-CO relation, the deviation from the sample's median at ${\rm L(IR)} \gtrsim 10^{12}\,L_{\odot}$ at $z\sim 0$ and ${\rm L(IR)} \gtrsim 10^{12.5}\,L_{\odot}$ at $z>1$ is noticeable (top panels of Figure~\ref{fig:lir}).
Both the PAH- and CO-to-IR luminosity ratios decrease significantly at high IR luminosities, also leading to sub-linear slopes for the best fit values (listed in Table~\ref{tab:fit}). This behaviour has been seen previously for both the PAH-to-IR \citep{pope13, stierwalt14, shipley16} and the CO-to-IR luminosities \citep{cortzen19, herrero19}. In local ULIRGs, the PAH luminosity is observed to decrease with increasing IR luminosity, possibly because the AGN emission starts to dominate the MIR emission \citep{desai07}. The observed trend of decreasing PAH luminosity with increasing IR luminosity occurs at higher IR luminosities at $z>1$ compared to that at $z\sim 0$, possibly because of the more extended star formation and lower surface density of ULIRGs at high redshifts \citep{rujopakarn13}, or relatively lower contribution from the AGN at fixed IR luminosity \citep[cf.][]{pope08a,pope08b,desai09}.
Similarly, for galaxies with efficient star formation activity, such as local ULIRGs and high-redshift SMGs with a high fraction of interacting galaxies, it is expected that the CO to IR luminosity ratio, which is proportional to gas depletion timescale, is lower than normal star forming galaxies (\citealt{daddi10a, pope13}, cf.\ \citealt{saintonge17, tacconi20}). 

In conclusion, from the presented compilation of galaxies at \(z \sim 0\) to 4, ranging from star-forming main sequence galaxies to ULIRGs and SMGs, the L(PAH)-L$^{\prime}$(CO) ratio is the only one that remains constant across the entire population. Both the L(PAH)-L(IR) and L$^{\prime}$(CO)-L(IR) relations are sub-linear, indicating a different coupling of the PAH and cold gas emission to the total dust luminosity in the ULIRG and HyLIRG regimes.

\section{Implications: PAH luminosity as a tracer of molecular gas mass} \label{sec:Mmol}

The linear and relatively tight and universal relation of L(PAH)-L$^{\prime}$(CO) intrigues the idea of adopting PAHs as a tracer of molecular gas mass, particularly in the era of sensitive MIR instruments onboard of {\em JWST}. 
Given an assumed CO luminosity to molecular gas mass conversion, $\alpha_{\rm CO}$, we propose a PAH 7.7{\um} luminosity to molecular gas mass conversion factor:
\begin{align}\label{eq:alphaPAH}
    \alpha_{\rm PAH_{7.7}} [\mathrm{M}_{\odot}/\mathrm{L}_{\odot}] &\equiv \frac{M_{\rm mol}}{L({\rm PAH}_{7.7})} = \frac{\alpha_{\rm CO}~L^{'}({\rm CO})}{L({\rm PAH}_{7.7})} = \frac{\alpha_{\rm CO}}{1.40 \pm 0.49},
\end{align}
which implies the molecular gas mass can be derived from the PAH luminosity as:
\begin{align} \label{eq:Mmol}
    M_{\rm mol}~[\mathrm{M}_{\rm \odot}] &= 3.08 \pm 1.08 \left(\frac{4.3}{\alpha_{\rm CO}}\right)~L({\rm PAH}_{7.7})~[\mathrm{L}_{\odot}].
\end{align}
Here, in Equation~\ref{eq:alphaPAH}, $1.40\pm0.49$ is the median ratio of PAH-to-CO luminosity ratio for the full sample reported in Table~\ref{tab:fit} with the MAD scatter (see Section~\ref{sec:results}).  Alternatively, the preferred $L({\rm PAH}_{7.7})$--$L'{\rm (CO)}$ relation from Table~\ref{tab:fit} can be inserted in Equation~\ref{eq:alphaPAH}. For the final value in Equation~\ref{eq:Mmol} we adopted a Milky-Way $\alpha_{\rm CO} = 4.3\,\mathrm{M}_{\odot}({\rm K~km~s^{-1}~pc^2})^{-1}$ \citep{bolatto13}. {The $\alpha_{\rm CO}$ conversion factor is known to vary among galaxies and galaxy types, and a Milky-Way like $\alpha_{\rm CO}$ has been found to be applicable to (massive, near-solar metallicity) star-forming galaxies in the local Universe and at higher redshifts \citep[e.g.][]{daddi10b}.  While a detailed discussion is beyond the scope of this work (see \citealt{bolatto13} for a review), we note that in extreme (nuclear) starburst conditions, such as ULIRGS and SMGs, a lower value of $\alpha_{\rm CO}$ may apply 
\citep[typically 0.8 is assumed;][though cf. 
\citealt{dunne22}]{downes98,papadopoulos12}, while in low-metallicity environments the conversion factor increases \citep{maloney88, israel97, wolfire10, sandstrom13}, which may become increasingly relevant for SFGs at high redshift \citep{boogaard21}.}

In the era of {\em JWST}, the luminosity of the high-equivalent width PAH features can be measured up to $z\sim 3$ using the MIRI imager, with modest ($\sim 10-30$ minute) integration times already reaching the LIRG regime and below, as demonstrated in this Letter \citep[see also][]{shen23,magnelli23,shivaei24,ronayne24}. This makes PAH observations far more accessible for large samples of galaxies at cosmic noon compared to CO observations with ALMA, which require significant time investments. This Letter shows that, owing to the tight and linear relationship between PAH 7.7{\um} and CO luminosity across a wide range of galaxy populations and redshifts, the proposed $\alpha_{\rm PAH_{7.7}}$ can be used as a valuable tool to explore the molecular gas content for large samples of galaxies.
This opens a new window for statistical studies of gas and star formation beyond our local Universe, not only in the ULIRG regime but also in typical populations of main-sequence galaxies.

\begin{table*}
\caption{PAH, CO, and IR luminosities of the main-sequence (ASPECS) sample.} \label{tab:data}           

\begin{tabular}{c c c c c c c c}
\hline\hline  
{ID\_3mm} & {RA} & {Dec} & {$z$} & {$\log(L(\mathrm{PAH}_{7.7})/\mathrm{L}_{\odot})$} & {$\log(L'(\rm CO)/{\rm K~km~s^{-1}~pc^2})$} & {$\log(L(\mathrm{IR})/L_{\odot})$} & {AGN} \\
\hline\hline
1  & 03:32:38.54 & -27:46:34.62 & 2.543 & 10.69 $\pm$ 0.03 & 10.51 $\pm$ 0.09 & 12.65 $\pm$ 0.03 & X-ray/IR \\
2  & 03:32:42.38 & -27:47:07.92 & 1.317 & 9.82 $\pm$ 0.16 & 10.15 $\pm$ 0.07 & 11.50 $\pm$ 0.04 & $-$ \\
3  & 03:32:41.02 & -27:46:31.56 & 2.454 & 10.30 $\pm$ 0.17 & 10.21 $\pm$ 0.09 & 12.01 $\pm$ -0.04 & $-$ \\
4  & 03:32:34.44 & -27:46:59.82 & 1.414 & 10.41 $\pm$ 0.06 & 10.49 $\pm$ 0.07 & 11.99 $\pm$ 0.02 & $-$ \\
5  & 03:32:39.76 & -27:46:11.58 & 1.551 & 10.49 $\pm$ 0.10 & 10.44 $\pm$ 0.07 & 12.01 $\pm$ 0.04 & X-ray \\
6  & 03:32:39.90 & -27:47:15.12 & 1.095 & 10.10 $\pm$ 0.04 & 10.01 $\pm$ 0.08 & 11.46 $\pm$ -0.01 & $-$ \\
7  & 03:32:43.53 & -27:46:39.47 & 2.696 & 10.70 $\pm$ 0.09 & 10.48 $\pm$ 0.15 & 12.44 $\pm$ -0.01 & $-$ \\
9  & 03:32:44.03 & -27:46:36.05 & 2.698 & 10.60 $\pm$ 0.15 & 10.11 $\pm$ 0.09 & 12.57 $\pm$ 0.07 & X-ray/IR \\
10 & 03:32:42.98 & -27:46:50.45 & 1.037 & 10.11 $\pm$ 0.06 & 10.05 $\pm$ 0.08 & 11.69 $\pm$ 0.03 & $-$ \\
11 & 03:32:39.80 & -27:46:53.70 & 1.096 & 9.44 $\pm$ 0.32 & 9.53 $\pm$ 0.10 & 10.91 $\pm$ 0.10 & $-$ \\
12 & 03:32:36.21 & -27:46:27.78 & 2.574 & 10.16 $\pm$ 0.13 & 9.77 $\pm$ 0.10 & 11.54 $\pm$ -0.04 & X-ray \\
14 & 03:32:34.84 & -27:46:40.74 & 1.098 & 9.97 $\pm$ 0.04 & 9.87 $\pm$ 0.09 & 11.46 $\pm$ -0.03 & $-$ \\
15 & 03:32:36.48 & -27:46:31.92 & 1.096 & 9.81 $\pm$ 0.05 & 9.65 $\pm$ 0.09 & 11.57 $\pm$ 0.01 & X-ray \\
16 & 03:32:39.92 & -27:46:07.44 & 1.294 & 9.75 $\pm$ 0.19 & 9.37 $\pm$ 0.08 & 11.19 $\pm$ -0.05 & $-$ \\
\hline 
\end{tabular}
Columns are, from left to right: ASPECS 3mm ID, Right Ascension, Declination, spectroscopic redshift, PAH 7.7{\um} luminosity, CO(1--0) luminosity, total IR luminosity, AGN flag (X-ray, IR, or no AGN). Note that 3mm.8 and 3mm.13 are not included; the former because it is largely behind a foreground galaxy, the latter because it is at $z>3$ and thus lacks the PAH coverage from MIRI.
\end{table*}

\begin{acknowledgements}
We thank the referee for a thorough and helpful report.
We warmly thank Tanio D{\'i}az Santos for providing feedback on the original manuscript.
We also thank the Heidelberg Joint Astrophysical Colloquium and the Cosmic Odysseys 2024 conference in Crete for providing a stimulating environment that facilitated this project.
Finally, we gratefully acknowledge the invaluable effort of our colleagues in the SMILES and ASPECS teams for generating the data that made this work possible.
I.S. acknowledges funding from Atracc{\' i}on de Talento Grant No.2022-T1/TIC-20472 of the Comunidad de Madrid, Spain. 
This work is based on observations made with the NASA/ESA/CSA \textit{James Webb} Space Telescope. The data were obtained from the Mikulski Archive for Space Telescopes at the Space Telescope Science Institute, which is operated by the Association of Universities for Research in Astronomy, Inc., under NASA contract NAS 5-03127 for JWST.  These observations are associated with program PID 1207, 1080, 1081, 1895, 1220, 1286, 1287, 1963.
Based on observations made with the NASA/ESA \textit{Hubble} Space Telescope, and obtained from the Hubble Legacy Archive, which is a collaboration between the Space Telescope Science Institute (STScI/NASA), the Space Telescope European Coordinating Facility (ST-ECF/ESAC/ESA) and the Canadian Astronomy Data Centre (CADC/NRC/CSA).
This paper makes use of the following ALMA data:
ADS/JAO.ALMA\#2016.1.00324.L.  ALMA is a partnership of ESO (representing its
member states), NSF (USA) and NINS (Japan), together with NRC (Canada), NSC and
ASIAA (Taiwan), and KASI (Republic of Korea), in cooperation with the Republic
of Chile. The Joint ALMA Observatory is operated by ESO, AUI/NRAO and NAOJ.
\end{acknowledgements} 

\bibliographystyle{aa}
\bibliography{literature}

\begin{thebibliography}{75}
\expandafter\ifx\csname natexlab\endcsname\relax\def\natexlab#1{#1}\fi

\bibitem[{{Alberts} {et~al.}(2024){Alberts}, {Lyu}, {Shivaei}, {Rieke},
  {Perez-Gonzalez}, {Bonventura}, {Zhu}, {Helton}, {Ji}, {Morrison},
  {Robertson}, {Stone}, {Sun}, {Williams}, \& {Willmer}}]{alberts24}
{Alberts}, S., {Lyu}, J., {Shivaei}, I., {et~al.} 2024, arXiv e-prints,
  arXiv:2405.15972

\bibitem[{{Armus} {et~al.}(2007){Armus}, {Charmandaris}, {Bernard-Salas},
  {Spoon}, {Marshall}, {Higdon}, {Desai}, {Teplitz}, {Hao}, {Devost}, {Brandl},
  {Wu}, {Sloan}, {Soifer}, {Houck}, \& {Herter}}]{armus07}
{Armus}, L., {Charmandaris}, V., {Bernard-Salas}, J., {et~al.} 2007, \apj, 656,
  148

\bibitem[{{Birkin} {et~al.}(2021){Birkin}, {Weiss}, {Wardlow}, {Smail},
  {Swinbank}, {Dudzevi{\v{c}}i{\={u}}t{\.{e}}}, {An}, {Ao}, {Chapman}, {Chen},
  {da Cunha}, {Dannerbauer}, {Gullberg}, {Hodge}, {Ikarashi}, {Ivison},
  {Matsuda}, {Stach}, {Walter}, {Wang}, \& {van der Werf}}]{birkin21}
{Birkin}, J.~E., {Weiss}, A., {Wardlow}, J.~L., {et~al.} 2021, \mnras, 501,
  3926

\bibitem[{{Bolatto} {et~al.}(2013){Bolatto}, {Wolfire}, \& {Leroy}}]{bolatto13}
{Bolatto}, A.~D., {Wolfire}, M., \& {Leroy}, A.~K. 2013, \araa, 51, 207

\bibitem[{{Boogaard} {et~al.}(2021){Boogaard}, {Bouwens}, {Riechers}, {van der
  Werf}, {Bacon}, {Matthee}, {Stefanon}, {Feltre}, {Maseda}, {Inami},
  {Aravena}, {Brinchmann}, {Carilli}, {Contini}, {Decarli},
  {Gonz{\'a}lez-L{\'o}pez}, {Nanayakkara}, \& {Walter}}]{boogaard21}
{Boogaard}, L.~A., {Bouwens}, R.~J., {Riechers}, D., {et~al.} 2021, \apj, 916,
  12

\bibitem[{{Boogaard} {et~al.}(2019){Boogaard}, {Decarli},
  {Gonz{\'a}lez-L{\'o}pez}, {van der Werf}, {Walter}, {Bouwens}, {Aravena},
  {Carilli}, {Bauer}, {Brinchmann}, {Contini}, {Cox}, {da Cunha}, {Daddi},
  {D{\'\i}az-Santos}, {Hodge}, {Inami}, {Ivison}, {Maseda}, {Matthee}, {Oesch},
  {Popping}, {Riechers}, {Schaye}, {Schouws}, {Smail}, {Weiss}, {Wisotzki},
  {Bacon}, {Cortes}, {Rix}, {Somerville}, {Swinbank}, \& {Wagg}}]{boogaard19}
{Boogaard}, L.~A., {Decarli}, R., {Gonz{\'a}lez-L{\'o}pez}, J., {et~al.} 2019,
  \apj, 882, 140

\bibitem[{{Boogaard} {et~al.}(2020){Boogaard}, {van der Werf}, {Weiss},
  {Popping}, {Decarli}, {Walter}, {Aravena}, {Bouwens}, {Riechers},
  {Gonz{\'a}lez-L{\'o}pez}, {Smail}, {Carilli}, {Kaasinen}, {Daddi}, {Cox},
  {D{\'\i}az-Santos}, {Inami}, {Cortes}, \& {Wagg}}]{boogaard20}
{Boogaard}, L.~A., {van der Werf}, P., {Weiss}, A., {et~al.} 2020, \apj, 902,
  109

\bibitem[{{Bothwell} {et~al.}(2013){Bothwell}, {Smail}, {Chapman}, {Genzel},
  {Ivison}, {Tacconi}, {Alaghband-Zadeh}, {Bertoldi}, {Blain}, {Casey}, {Cox},
  {Greve}, {Lutz}, {Neri}, {Omont}, \& {Swinbank}}]{bothwell13}
{Bothwell}, M.~S., {Smail}, I., {Chapman}, S.~C., {et~al.} 2013, \mnras, 429,
  3047

\bibitem[{{Brandl} {et~al.}(2006){Brandl}, {Bernard-Salas}, {Spoon}, {Devost},
  {Sloan}, {Guilles}, {Wu}, {Houck}, {Weedman}, {Armus}, {Appleton}, {Soifer},
  {Charmandaris}, {Hao}, {Higdon}, {Marshall}, \& {Herter}}]{brandl06}
{Brandl}, B.~R., {Bernard-Salas}, J., {Spoon}, H.~W.~W., {et~al.} 2006, \apj,
  653, 1129

\bibitem[{{Calzetti}(2011)}]{calzetti11}
{Calzetti}, D. 2011, in EAS Publications Series, Vol.~46, EAS Publications
  Series, ed. C.~{Joblin} \& A.~G.~G.~M. {Tielens}, 133--141

\bibitem[{{Carilli} \& {Walter}(2013)}]{carilli13}
{Carilli}, C.~L. \& {Walter}, F. 2013, \araa, 51, 105

\bibitem[{{Chown} {et~al.}(2021){Chown}, {Li}, {Parker}, {Wilson}, {Li}, \&
  {Gao}}]{chown21}
{Chown}, R., {Li}, C., {Parker}, L., {et~al.} 2021, \mnras, 500, 1261

\bibitem[{{Cortzen} {et~al.}(2019){Cortzen}, {Garrett}, {Magdis}, {Rigopoulou},
  {Valentino}, {Pereira-Santaella}, {Combes}, {Alonso-Herrero}, {Toft},
  {Daddi}, {Elbaz}, {G{\'o}mez-Guijarro}, {Stockmann}, {Huang}, \&
  {Kramer}}]{cortzen19}
{Cortzen}, I., {Garrett}, J., {Magdis}, G., {et~al.} 2019, \mnras, 482, 1618

\bibitem[{{Daddi} {et~al.}(2010{\natexlab{a}}){Daddi}, {Bournaud}, {Walter},
  {Dannerbauer}, {Carilli}, {Dickinson}, {Elbaz}, {Morrison}, {Riechers},
  {Onodera}, {Salmi}, {Krips}, \& {Stern}}]{daddi10a}
{Daddi}, E., {Bournaud}, F., {Walter}, F., {et~al.} 2010{\natexlab{a}}, \apj,
  713, 686

\bibitem[{{Daddi} {et~al.}(2010{\natexlab{b}}){Daddi}, {Elbaz}, {Walter},
  {Bournaud}, {Salmi}, {Carilli}, {Dannerbauer}, {Dickinson}, {Monaco}, \&
  {Riechers}}]{daddi10b}
{Daddi}, E., {Elbaz}, D., {Walter}, F., {et~al.} 2010{\natexlab{b}}, \apjl,
  714, L118

\bibitem[{{Decarli} {et~al.}(2020){Decarli}, {Aravena}, {Boogaard}, {Carilli},
  {Gonz{\'a}lez-L{\'o}pez}, {Walter}, {Cortes}, {Cox}, {da Cunha}, {Daddi},
  {D{\'\i}az-Santos}, {Hodge}, {Inami}, {Neeleman}, {Novak}, {Oesch},
  {Popping}, {Riechers}, {Smail}, {Uzgil}, {van der Werf}, {Wagg}, \&
  {Weiss}}]{decarli20}
{Decarli}, R., {Aravena}, M., {Boogaard}, L., {et~al.} 2020, \apj, 902, 110

\bibitem[{{Decarli} {et~al.}(2019){Decarli}, {Walter},
  {G{\'o}nzalez-L{\'o}pez}, {Aravena}, {Boogaard}, {Carilli}, {Cox}, {Daddi},
  {Popping}, {Riechers}, {Uzgil}, {Weiss}, {Assef}, {Bacon}, {Bauer},
  {Bertoldi}, {Bouwens}, {Contini}, {Cortes}, {da Cunha}, {D{\'\i}az-Santos},
  {Elbaz}, {Inami}, {Hodge}, {Ivison}, {Le F{\`e}vre}, {Magnelli}, {Novak},
  {Oesch}, {Rix}, {Sargent}, {Smail}, {Swinbank}, {Somerville}, {van der Werf},
  {Wagg}, \& {Wisotzki}}]{decarli19}
{Decarli}, R., {Walter}, F., {G{\'o}nzalez-L{\'o}pez}, J., {et~al.} 2019, \apj,
  882, 138

\bibitem[{{Desai} {et~al.}(2007){Desai}, {Armus}, {Spoon}, {Charmandaris},
  {Bernard-Salas}, {Brandl}, {Farrah}, {Soifer}, {Teplitz}, {Ogle}, {Devost},
  {Higdon}, {Marshall}, \& {Houck}}]{desai07}
{Desai}, V., {Armus}, L., {Spoon}, H.~W.~W., {et~al.} 2007, \apj, 669, 810

\bibitem[{{Desai} {et~al.}(2009){Desai}, {Soifer}, {Dey}, {Le Floc'h}, {Armus},
  {Brand}, {Brown}, {Brodwin}, {Jannuzi}, {Houck}, {Weedman}, {Ashby},
  {Gonzalez}, {Huang}, {Smith}, {Teplitz}, {Willner}, \& {Melbourne}}]{desai09}
{Desai}, V., {Soifer}, B.~T., {Dey}, A., {et~al.} 2009, \apj, 700, 1190

\bibitem[{{Donnan} {et~al.}(2024){Donnan}, {Garc{\'\i}a-Bernete}, {Rigopoulou},
  {Pereira-Santaella}, {Roche}, \& {Alonso-Herrero}}]{donnan24}
{Donnan}, F.~R., {Garc{\'\i}a-Bernete}, I., {Rigopoulou}, D., {et~al.} 2024,
  \mnras, 529, 1386

\bibitem[{{Downes} \& {Solomon}(1998)}]{downes98}
{Downes}, D. \& {Solomon}, P.~M. 1998, \apj, 507, 615

\bibitem[{{Draine} \& {Li}(2001)}]{draine01}
{Draine}, B.~T. \& {Li}, A. 2001, ApJ, 551, 807

\bibitem[{{Draine} \& {Li}(2007)}]{DL07}
{Draine}, B.~T. \& {Li}, A. 2007, \apj, 657, 810

\bibitem[{{Draine} {et~al.}(2021){Draine}, {Li}, {Hensley}, {Hunt},
  {Sandstrom}, \& {Smith}}]{draine21}
{Draine}, B.~T., {Li}, A., {Hensley}, B.~S., {et~al.} 2021, \apj, 917, 3

\bibitem[{{Dunne} {et~al.}(2022){Dunne}, {Maddox}, {Papadopoulos}, {Ivison}, \&
  {Gomez}}]{dunne22}
{Dunne}, L., {Maddox}, S.~J., {Papadopoulos}, P.~P., {Ivison}, R.~J., \&
  {Gomez}, H.~L. 2022, \mnras, 517, 962

\bibitem[{{Elbaz} {et~al.}(2011){Elbaz}, {Dickinson}, {Hwang},
  {D{\'\i}az-Santos}, {Magdis}, {Magnelli}, {Le Borgne}, {Galliano},
  {Pannella}, {Chanial}, {Armus}, {Charmandaris}, {Daddi}, {Aussel}, {Popesso},
  {Kartaltepe}, {Altieri}, {Valtchanov}, {Coia}, {Dannerbauer}, {Dasyra},
  {Leiton}, {Mazzarella}, {Alexander}, {Buat}, {Burgarella}, {Chary}, {Gilli},
  {Ivison}, {Juneau}, {Le Floc'h}, {Lutz}, {Morrison}, {Mullaney}, {Murphy},
  {Pope}, {Scott}, {Brodwin}, {Calzetti}, {Cesarsky}, {Charlot}, {Dole},
  {Eisenhardt}, {Ferguson}, {F{\"o}rster Schreiber}, {Frayer}, {Giavalisco},
  {Huynh}, {Koekemoer}, {Papovich}, {Reddy}, {Surace}, {Teplitz}, {Yun}, \&
  {Wilson}}]{elbaz11}
{Elbaz}, D., {Dickinson}, M., {Hwang}, H.~S., {et~al.} 2011, \aap, 533, A119

\bibitem[{{Gao} {et~al.}(2022){Gao}, {Tan}, {Gao}, {Fang}, {Chown}, {Jiao}, \&
  {Luo}}]{gao22}
{Gao}, Y., {Tan}, Q.-H., {Gao}, Y., {et~al.} 2022, \apj, 940, 133

\bibitem[{{Garc{\'\i}a-Bernete} {et~al.}(2022){Garc{\'\i}a-Bernete},
  {Rigopoulou}, {Alonso-Herrero}, {Donnan}, {Roche}, {Pereira-Santaella},
  {Labiano}, {Peralta de Arriba}, {Izumi}, {Ramos Almeida}, {Shimizu},
  {H{\"o}nig}, {Garc{\'\i}a-Burillo}, {Rosario}, {Ward}, {Bellocchi}, {Hicks},
  {Fuller}, \& {Packham}}]{garcia-bernete22}
{Garc{\'\i}a-Bernete}, I., {Rigopoulou}, D., {Alonso-Herrero}, A., {et~al.}
  2022, \aap, 666, L5

\bibitem[{{Gonz{\'a}lez-L{\'o}pez} {et~al.}(2019){Gonz{\'a}lez-L{\'o}pez},
  {Decarli}, {Pavesi}, {Walter}, {Aravena}, {Carilli}, {Boogaard}, {Popping},
  {Weiss}, {Assef}, {Bauer}, {Bertoldi}, {Bouwens}, {Contini}, {Cortes}, {Cox},
  {da Cunha}, {Daddi}, {D{\'\i}az-Santos}, {Inami}, {Hodge}, {Ivison}, {Le
  F{\`e}vre}, {Magnelli}, {Oesch}, {Riechers}, {Rix}, {Smail}, {Swinbank},
  {Somerville}, {Uzgil}, \& {van der Werf}}]{gonzalez-lopez19}
{Gonz{\'a}lez-L{\'o}pez}, J., {Decarli}, R., {Pavesi}, R., {et~al.} 2019, \apj,
  882, 139

\bibitem[{{Grogin} {et~al.}(2011){Grogin}, {Kocevski}, {Faber}, {Ferguson},
  {Koekemoer}, {Riess}, {Acquaviva}, {Alexander}, {Almaini}, {Ashby}, {Barden},
  {Bell}, {Bournaud}, {Brown}, {Caputi}, {Casertano}, {Cassata}, {Castellano},
  {Challis}, {Chary}, {Cheung}, {Cirasuolo}, {Conselice}, {Roshan Cooray},
  {Croton}, {Daddi}, {Dahlen}, {Dav{\'e}}, {de Mello}, {Dekel}, {Dickinson},
  {Dolch}, {Donley}, {Dunlop}, {Dutton}, {Elbaz}, {Fazio}, {Filippenko},
  {Finkelstein}, {Fontana}, {Gardner}, {Garnavich}, {Gawiser}, {Giavalisco},
  {Grazian}, {Guo}, {Hathi}, {H{\"a}ussler}, {Hopkins}, {Huang}, {Huang},
  {Jha}, {Kartaltepe}, {Kirshner}, {Koo}, {Lai}, {Lee}, {Li}, {Lotz}, {Lucas},
  {Madau}, {McCarthy}, {McGrath}, {McIntosh}, {McLure}, {Mobasher},
  {Moustakas}, {Mozena}, {Nandra}, {Newman}, {Niemi}, {Noeske}, {Papovich},
  {Pentericci}, {Pope}, {Primack}, {Rajan}, {Ravindranath}, {Reddy}, {Renzini},
  {Rix}, {Robaina}, {Rodney}, {Rosario}, {Rosati}, {Salimbeni}, {Scarlata},
  {Siana}, {Simard}, {Smidt}, {Somerville}, {Spinrad}, {Straughn}, {Strolger},
  {Telford}, {Teplitz}, {Trump}, {van der Wel}, {Villforth}, {Wechsler},
  {Weiner}, {Wiklind}, {Wild}, {Wilson}, {Wuyts}, {Yan}, \& {Yun}}]{grogin11}
{Grogin}, N.~A., {Kocevski}, D.~D., {Faber}, S.~M., {et~al.} 2011, \apjs, 197,
  35

\bibitem[{{Haas} {et~al.}(2002){Haas}, {Klaas}, \& {Bianchi}}]{haas02}
{Haas}, M., {Klaas}, U., \& {Bianchi}, S. 2002, \aap, 385, L23

\bibitem[{{Herrero-Illana} {et~al.}(2019){Herrero-Illana}, {Privon}, {Evans},
  {D{\'\i}az-Santos}, {P{\'e}rez-Torres}, {U}, {Alberdi}, {Iwasawa}, {Armus},
  {Aalto}, {Mazzarella}, {Chu}, {Sanders}, {Barcos-Mu{\~n}oz}, {Charmandaris},
  {Linden}, {Yoon}, {Frayer}, {Inami}, {Kim}, {Borish}, {Conway}, {Murphy},
  {Song}, {Stierwalt}, \& {Surace}}]{herrero19}
{Herrero-Illana}, R., {Privon}, G.~C., {Evans}, A.~S., {et~al.} 2019, \aap,
  628, A71

\bibitem[{{Israel}(1997)}]{israel97}
{Israel}, F.~P. 1997, \aap, 328, 471

\bibitem[{{Johnson} {et~al.}(2021){Johnson}, {Leja}, {Conroy}, \&
  {Speagle}}]{johnson21}
{Johnson}, B.~D., {Leja}, J., {Conroy}, C., \& {Speagle}, J.~S. 2021, \apjs,
  254, 22

\bibitem[{{Kennicutt} \& {De Los Reyes}(2021)}]{kennicutt21}
{Kennicutt}, Robert~C., J. \& {De Los Reyes}, M. A.~C. 2021, \apj, 908, 61

\bibitem[{{Kennicutt} \& {Evans}(2012)}]{kennicutt12}
{Kennicutt}, R.~C. \& {Evans}, N.~J. 2012, \araa, 50, 531

\bibitem[{{Kirkpatrick} {et~al.}(2014){Kirkpatrick}, {Calzetti}, {Kennicutt},
  {Galametz}, {Gordon}, {Groves}, {Hunt}, {Dale}, {Hinz}, \&
  {Tabatabaei}}]{kirkpatrick14}
{Kirkpatrick}, A., {Calzetti}, D., {Kennicutt}, R., {et~al.} 2014, \apj, 789,
  130

\bibitem[{{Koekemoer} {et~al.}(2011){Koekemoer}, {Faber}, {Ferguson}, {Grogin},
  {Kocevski}, {Koo}, {Lai}, {Lotz}, {Lucas}, {McGrath}, {Ogaz}, {Rajan},
  {Riess}, {Rodney}, {Strolger}, {Casertano}, {Castellano}, {Dahlen},
  {Dickinson}, {Dolch}, {Fontana}, {Giavalisco}, {Grazian}, {Guo}, {Hathi},
  {Huang}, {van der Wel}, {Yan}, {Acquaviva}, {Alexander}, {Almaini}, {Ashby},
  {Barden}, {Bell}, {Bournaud}, {Brown}, {Caputi}, {Cassata}, {Challis},
  {Chary}, {Cheung}, {Cirasuolo}, {Conselice}, {Roshan Cooray}, {Croton},
  {Daddi}, {Dav{\'e}}, {de Mello}, {de Ravel}, {Dekel}, {Donley}, {Dunlop},
  {Dutton}, {Elbaz}, {Fazio}, {Filippenko}, {Finkelstein}, {Frazer}, {Gardner},
  {Garnavich}, {Gawiser}, {Gruetzbauch}, {Hartley}, {H{\"a}ussler},
  {Herrington}, {Hopkins}, {Huang}, {Jha}, {Johnson}, {Kartaltepe},
  {Khostovan}, {Kirshner}, {Lani}, {Lee}, {Li}, {Madau}, {McCarthy},
  {McIntosh}, {McLure}, {McPartland}, {Mobasher}, {Moreira}, {Mortlock},
  {Moustakas}, {Mozena}, {Nandra}, {Newman}, {Nielsen}, {Niemi}, {Noeske},
  {Papovich}, {Pentericci}, {Pope}, {Primack}, {Ravindranath}, {Reddy},
  {Renzini}, {Rix}, {Robaina}, {Rosario}, {Rosati}, {Salimbeni}, {Scarlata},
  {Siana}, {Simard}, {Smidt}, {Snyder}, {Somerville}, {Spinrad}, {Straughn},
  {Telford}, {Teplitz}, {Trump}, {Vargas}, {Villforth}, {Wagner}, {Wandro},
  {Wechsler}, {Weiner}, {Wiklind}, {Wild}, {Wilson}, {Wuyts}, \&
  {Yun}}]{koekemoer11}
{Koekemoer}, A.~M., {Faber}, S.~M., {Ferguson}, H.~C., {et~al.} 2011, \apjs,
  197, 36

\bibitem[{{Leroy} {et~al.}(2023{\natexlab{a}}){Leroy}, {Bolatto}, {Sandstrom},
  {Rosolowsky}, {Barnes}, {Bigiel}, {Boquien}, {den Brok}, {Cao}, {Chastenet},
  {Chevance}, {Chiang}, {Chown}, {Colombo}, {Ellison}, {Emsellem}, {Grasha},
  {Henshaw}, {Hughes}, {Klessen}, {Koch}, {Kim}, {Kreckel}, {Kruijssen},
  {Larson}, {Lee}, {Levy}, {Lin}, {Liu}, {Meidt}, {Pety}, {Querejeta}, {Rubio},
  {Saito}, {Salim}, {Schinnerer}, {Sormani}, {Sun}, {Thilker}, {Usero},
  {Vogel}, {Watkins}, {Whitcomb}, {Williams}, \& {Wilson}}]{leroy23a}
{Leroy}, A.~K., {Bolatto}, A.~D., {Sandstrom}, K., {et~al.} 2023{\natexlab{a}},
  \apjl, 944, L10

\bibitem[{{Leroy} {et~al.}(2023{\natexlab{b}}){Leroy}, {Sandstrom},
  {Rosolowsky}, {Belfiore}, {Bolatto}, {Cao}, {Koch}, {Schinnerer}, {Barnes},
  {Be{\v{s}}li{\'c}}, {Bigiel}, {Blanc}, {Chastenet}, {Chen}, {Chevance},
  {Chown}, {Congiu}, {Dale}, {Egorov}, {Emsellem}, {Eibensteiner}, {Faesi},
  {Glover}, {Grasha}, {Groves}, {Hassani}, {Henshaw}, {Hughes},
  {Jim{\'e}nez-Donaire}, {Kim}, {Klessen}, {Kreckel}, {Kruijssen}, {Larson},
  {Lee}, {Levy}, {Liu}, {Lopez}, {Meidt}, {Murphy}, {Neumann}, {Pessa}, {Pety},
  {Saito}, {Sardone}, {Sun}, {Thilker}, {Usero}, {Watkins}, {Whitcomb}, \&
  {Williams}}]{leroy23b}
{Leroy}, A.~K., {Sandstrom}, K., {Rosolowsky}, E., {et~al.} 2023{\natexlab{b}},
  \apjl, 944, L9

\bibitem[{{Li} \& {Draine}(2002)}]{li02}
{Li}, A. \& {Draine}, B.~T. 2002, \apj, 572, 232

\bibitem[{{Luo} {et~al.}(2017){Luo}, {Brandt}, {Xue}, {Lehmer}, {Alexander},
  {Bauer}, {Vito}, {Yang}, {Basu-Zych}, {Comastri}, {Gilli}, {Gu},
  {Hornschemeier}, {Koekemoer}, {Liu}, {Mainieri}, {Paolillo}, {Ranalli},
  {Rosati}, {Schneider}, {Shemmer}, {Smail}, {Sun}, {Tozzi}, {Vignali}, \&
  {Wang}}]{luo17}
{Luo}, B., {Brandt}, W.~N., {Xue}, Y.~Q., {et~al.} 2017, \apjs, 228, 2

\bibitem[{{Magnelli} {et~al.}(2023){Magnelli}, {G{\'o}mez-Guijarro}, {Elbaz},
  {Daddi}, {Papovich}, {Shen}, {Arrabal Haro}, {Bagley}, {Bell}, {Buat},
  {Costantin}, {Dickinson}, {Finkelstein}, {Gardner}, {Jim{\'e}nez-Andrade},
  {Kartaltepe}, {Koekemoer}, {Lyu}, {P{\'e}rez-Gonz{\'a}lez}, {Pirzkal},
  {Tacchella}, {de la Vega}, {Wuyts}, {Yang}, {Yung}, \& {Zavala}}]{magnelli23}
{Magnelli}, B., {G{\'o}mez-Guijarro}, C., {Elbaz}, D., {et~al.} 2023, \aap,
  678, A83

\bibitem[{{Maloney} \& {Black}(1988)}]{maloney88}
{Maloney}, P. \& {Black}, J.~H. 1988, \apj, 325, 389

\bibitem[{{McKee} \& {Ostriker}(2007)}]{mckee07}
{McKee}, C.~F. \& {Ostriker}, E.~C. 2007, \araa, 45, 565

\bibitem[{{Papadopoulos} {et~al.}(2012){Papadopoulos}, {van der Werf},
  {Xilouris}, {Isaak}, \& {Gao}}]{papadopoulos12}
{Papadopoulos}, P.~P., {van der Werf}, P., {Xilouris}, E., {Isaak}, K.~G., \&
  {Gao}, Y. 2012, \apj, 751, 10

\bibitem[{{Pope} {et~al.}(2008{\natexlab{a}}){Pope}, {Bussmann}, {Dey},
  {Meger}, {Alexander}, {Brodwin}, {Chary}, {Dickinson}, {Frayer}, {Greve},
  {Huynh}, {Lin}, {Morrison}, {Scott}, \& {Yan}}]{pope08b}
{Pope}, A., {Bussmann}, R.~S., {Dey}, A., {et~al.} 2008{\natexlab{a}}, \apj,
  689, 127

\bibitem[{{Pope} {et~al.}(2008{\natexlab{b}}){Pope}, {Chary}, {Alexander},
  {Armus}, {Dickinson}, {Elbaz}, {Frayer}, {Scott}, \& {Teplitz}}]{pope08a}
{Pope}, A., {Chary}, R.-R., {Alexander}, D.~M., {et~al.} 2008{\natexlab{b}},
  \apj, 675, 1171

\bibitem[{{Pope} {et~al.}(2013){Pope}, {Wagg}, {Frayer}, {Armus}, {Chary},
  {Daddi}, {Desai}, {Dickinson}, {Elbaz}, {Gabor}, \& {Kirkpatrick}}]{pope13}
{Pope}, A., {Wagg}, J., {Frayer}, D., {et~al.} 2013, \apj, 772, 92

\bibitem[{{Riechers} {et~al.}(2020){Riechers}, {Boogaard}, {Decarli},
  {Gonz{\'a}lez-L{\'o}pez}, {Smail}, {Walter}, {Aravena}, {Carilli}, {Cortes},
  {Cox}, {D{\'\i}az-Santos}, {Hodge}, {Inami}, {Ivison}, {Kaasinen}, {Wagg},
  {Wei{\ss}}, \& {van der Werf}}]{riechers20}
{Riechers}, D.~A., {Boogaard}, L.~A., {Decarli}, R., {et~al.} 2020, \apjl, 896,
  L21

\bibitem[{{Rieke} {et~al.}(2024){Rieke}, {Alberts}, {Shivaei}, {Lyu},
  {Willmer}, {Perez-Gonzalez}, \& {Williams}}]{rieke24}
{Rieke}, G., {Alberts}, S., {Shivaei}, I., {et~al.} 2024, arXiv e-prints,
  arXiv:2406.03518

\bibitem[{{Rieke} {et~al.}(2015){Rieke}, {Wright}, {B{\"o}ker}, {Bouwman},
  {Colina}, {Glasse}, {Gordon}, {Greene}, {G{\"u}del}, {Henning}, {Justtanont},
  {Lagage}, {Meixner}, {N{\o}rgaard-Nielsen}, {Ray}, {Ressler}, {van Dishoeck},
  \& {Waelkens}}]{rieke15}
{Rieke}, G.~H., {Wright}, G.~S., {B{\"o}ker}, T., {et~al.} 2015, \pasp, 127,
  584

\bibitem[{{Rieke} {et~al.}(2023){Rieke}, {Robertson}, {Tacchella}, {Hainline},
  {Johnson}, {Hausen}, {Ji}, {Willmer}, {Eisenstein}, {Pusk{\'a}s}, {Alberts},
  {Arribas}, {Baker}, {Baum}, {Bhatawdekar}, {Bonaventura}, {Boyett}, {Bunker},
  {Cameron}, {Carniani}, {Charlot}, {Chevallard}, {Chen}, {Curti},
  {Curtis-Lake}, {Danhaive}, {DeCoursey}, {Dressler}, {Egami}, {Endsley},
  {Helton}, {Hviding}, {Kumari}, {Looser}, {Lyu}, {Maiolino}, {Maseda},
  {Nelson}, {Rieke}, {Rix}, {Sandles}, {Saxena}, {Sharpe}, {Shivaei},
  {Skarbinski}, {Smit}, {Stark}, {Stone}, {Suess}, {Sun}, {Topping},
  {{\"U}bler}, {Villanueva}, {Wallace}, {Williams}, {Willott}, {Whitler},
  {Witstok}, \& {Woodrum}}]{rieke23}
{Rieke}, M.~J., {Robertson}, B., {Tacchella}, S., {et~al.} 2023, \apjs, 269, 16

\bibitem[{{Robotham} \& {Obreschkow}(2015)}]{robotham2015}
{Robotham}, A.~S.~G. \& {Obreschkow}, D. 2015, \pasa, 32, e033

\bibitem[{{Ronayne} {et~al.}(2024){Ronayne}, {Papovich}, {Yang}, {Shen},
  {Dickinson}, {Kennicutt}, {Alavi}, {Arrabal Haro}, {Bagley}, {Burgarella},
  {Le Bail}, {Bell}, {Cleri}, {Cole}, {Costantin}, {de la Vega}, {Daddi},
  {Elbaz}, {Finkelstein}, {Grogin}, {Holwerda}, {Kartaltepe}, {Kirkpatrick},
  {Koekemoer}, {Lucas}, {Magnelli}, {Mobasher}, {P{\'e}rez-Gonz{\'a}lez},
  {Prichard}, {Rafelski}, {Rodighiero}, {Sunnquist}, {Teplitz}, {Wang},
  {Windhorst}, \& {Yung}}]{ronayne24}
{Ronayne}, K., {Papovich}, C., {Yang}, G., {et~al.} 2024, \apj, 970, 61

\bibitem[{{Rujopakarn} {et~al.}(2013){Rujopakarn}, {Rieke}, {Weiner},
  {P{\'e}rez-Gonz{\'a}lez}, {Rex}, {Walth}, \& {Kartaltepe}}]{rujopakarn13}
{Rujopakarn}, W., {Rieke}, G.~H., {Weiner}, B.~J., {et~al.} 2013, \apj, 767, 73

\bibitem[{{Saintonge} {et~al.}(2017){Saintonge}, {Catinella}, {Tacconi},
  {Kauffmann}, {Genzel}, {Cortese}, {Dav{\'e}}, {Fletcher},
  {Graci{\'a}-Carpio}, {Kramer}, {Heckman}, {Janowiecki}, {Lutz}, {Rosario},
  {Schiminovich}, {Schuster}, {Wang}, {Wuyts}, {Borthakur}, {Lamperti}, \&
  {Roberts-Borsani}}]{saintonge17}
{Saintonge}, A., {Catinella}, B., {Tacconi}, L.~J., {et~al.} 2017, \apjs, 233,
  22

\bibitem[{{Sandstrom} {et~al.}(2013){Sandstrom}, {Leroy}, {Walter}, {Bolatto},
  {Croxall}, {Draine}, {Wilson}, {Wolfire}, {Calzetti}, {Kennicutt}, {Aniano},
  {Donovan Meyer}, {Usero}, {Bigiel}, {Brinks}, {de Blok}, {Crocker}, {Dale},
  {Engelbracht}, {Galametz}, {Groves}, {Hunt}, {Koda}, {Kreckel}, {Linz},
  {Meidt}, {Pellegrini}, {Rix}, {Roussel}, {Schinnerer}, {Schruba}, {Schuster},
  {Skibba}, {van der Laan}, {Appleton}, {Armus}, {Brandl}, {Gordon}, {Hinz},
  {Krause}, {Montiel}, {Sauvage}, {Schmiedeke}, {Smith}, \&
  {Vigroux}}]{sandstrom13}
{Sandstrom}, K.~M., {Leroy}, A.~K., {Walter}, F., {et~al.} 2013, \apj, 777, 5

\bibitem[{{Sellgren}(2001)}]{sellgren01}
{Sellgren}, K. 2001, Spectrochimica Acta Part A: Molecular Spectroscopy, 57,
  627

\bibitem[{{Shen} {et~al.}(2023){Shen}, {Papovich}, {Yang}, {Matharu}, {Wang},
  {Magnelli}, {Elbaz}, {Jogee}, {Alavi}, {Arrabal Haro}, {Backhaus}, {Bagley},
  {Bell}, {Bisigello}, {Calabr{\`o}}, {Cooper}, {Costantin}, {Daddi},
  {Dickinson}, {Finkelstein}, {Fujimoto}, {Giavalisco}, {Grogin}, {Guo},
  {Holwerda}, {Kartaltepe}, {Koekemoer}, {Kurczynski}, {Lucas},
  {P{\'e}rez-Gonz{\'a}lez}, {Pirzkal}, {Prichard}, {Rafelski}, {Ronayne},
  {Simons}, {Sunnquist}, {Teplitz}, {Trump}, {Weiner}, {Windhorst}, \&
  {Yung}}]{shen23}
{Shen}, L., {Papovich}, C., {Yang}, G., {et~al.} 2023, \apj, 950, 7

\bibitem[{{Shipley} {et~al.}(2016){Shipley}, {Papovich}, {Rieke}, {Brown}, \&
  {Moustakas}}]{shipley16}
{Shipley}, H.~V., {Papovich}, C., {Rieke}, G.~H., {Brown}, M. J.~I., \&
  {Moustakas}, J. 2016, \apj, 818, 60

\bibitem[{{Shivaei} {et~al.}(2024){Shivaei}, {Alberts}, {Florian}, {Rieke},
  {Wuyts}, {Bodansky}, {Bunker}, {Cameron}, {Curti}, {D'Eugenio},
  {Dudzeviciute}, {Kramarenko}, {Ji}, {Johnson}, {Lyu}, {Matthee}, {Morrison},
  {Naidu}, {Reddy}, {Robertson}, {P{\'e}rez-Gonz{\'a}lez}, {Sun}, {Tacchella},
  {Whitaker}, {Williams}, {Willmer}, {Witstok}, {Xiao}, \& {Zhu}}]{shivaei24}
{Shivaei}, I., {Alberts}, S., {Florian}, M., {et~al.} 2024, arXiv e-prints,
  arXiv:2402.07989

\bibitem[{{Shivaei} {et~al.}(2015){Shivaei}, {Reddy}, {Shapley}, {Kriek},
  {Siana}, {Mobasher}, {Coil}, {Freeman}, {Sanders}, {Price}, {de Groot}, \&
  {Azadi}}]{shivaei15}
{Shivaei}, I., {Reddy}, N.~A., {Shapley}, A.~E., {et~al.} 2015, \apj, 815, 98

\bibitem[{{Shivaei} {et~al.}(2017){Shivaei}, {Reddy}, {Shapley}, {Siana},
  {Kriek}, {Mobasher}, {Coil}, {Freeman}, {Sanders}, {Price}, {Azadi}, \&
  {Zick}}]{shivaei17}
{Shivaei}, I., {Reddy}, N.~A., {Shapley}, A.~E., {et~al.} 2017, \apj, 837, 157

\bibitem[{Smith {et~al.}(2007)Smith, Draine, Dale, Moustakas, Kennicutt, Helou,
  Armus, Roussel, Sheth, Bendo, Buckalew, Calzetti, Engelbracht, Gordon,
  Hollenbach, Li, Malhotra, Murphy, \& Walter}]{smith07}
Smith, J. D.~T., Draine, B.~T., Dale, D.~A., {et~al.} 2007, ApJ, 656, 770

\bibitem[{{Stierwalt} {et~al.}(2014){Stierwalt}, {Armus}, {Charmandaris},
  {Diaz-Santos}, {Marshall}, {Evans}, {Haan}, {Howell}, {Iwasawa}, {Kim},
  {Murphy}, {Rich}, {Spoon}, {Inami}, {Petric}, \& {U}}]{stierwalt14}
{Stierwalt}, S., {Armus}, L., {Charmandaris}, V., {et~al.} 2014, \apj, 790, 124

\bibitem[{{Tacconi} {et~al.}(2020){Tacconi}, {Genzel}, \&
  {Sternberg}}]{tacconi20}
{Tacconi}, L.~J., {Genzel}, R., \& {Sternberg}, A. 2020, \araa, 58, 157

\bibitem[{{Tielens}(2008)}]{tielens08}
{Tielens}, A.~G.~G.~M. 2008, \araa, 46, 289

\bibitem[{{Whitaker} {et~al.}(2017){Whitaker}, {Pope}, {Cybulski}, {Casey},
  {Popping}, \& {Yun}}]{whitaker17}
{Whitaker}, K.~E., {Pope}, A., {Cybulski}, R., {et~al.} 2017, \apj, 850, 208

\bibitem[{{Whitcomb} {et~al.}(2023){Whitcomb}, {Sandstrom}, {Leroy}, \&
  {Smith}}]{whitcomb23}
{Whitcomb}, C.~M., {Sandstrom}, K., {Leroy}, A., \& {Smith}, J. D.~T. 2023,
  \apj, 948, 88

\bibitem[{{Wolfire} {et~al.}(2010){Wolfire}, {Hollenbach}, \&
  {McKee}}]{wolfire10}
{Wolfire}, M.~G., {Hollenbach}, D., \& {McKee}, C.~F. 2010, \apj, 716, 1191

\bibitem[{{Wright} {et~al.}(2023){Wright}, {Rieke}, {Glasse}, {Ressler},
  {Garc{\'\i}a Mar{\'\i}n}, {Aguilar}, {Alberts}, {{\'A}lvarez-M{\'a}rquez},
  {Argyriou}, {Banks}, {Baudoz}, {Boccaletti}, {Bouchet}, {Bouwman}, {Brandl},
  {Breda}, {Bright}, {Cale}, {Colina}, {Cossou}, {Coulais}, {Cracraft}, {De
  Meester}, {Dicken}, {Engesser}, {Etxaluze}, {Fox}, {Friedman}, {Fu},
  {Gasman}, {G{\'a}sp{\'a}r}, {Gastaud}, {Geers}, {Glauser}, {Gordon},
  {Greene}, {Greve}, {Grundy}, {G{\"u}del}, {Guillard}, {Haderlein},
  {Hashimoto}, {Henning}, {Hines}, {Holler}, {Detre}, {Jahromi}, {James},
  {Jones}, {Justtanont}, {Kavanagh}, {Kendrew}, {Klaassen}, {Krause},
  {Labiano}, {Lagage}, {Lambros}, {Larson}, {Law}, {Lee}, {Libralato}, {Lorenzo
  Alverez}, {Meixner}, {Morrison}, {Mueller}, {Murray}, {Mycroft}, {Myers},
  {Nayak}, {Naylor}, {Nickson}, {Noriega-Crespo}, {{\"O}stlin}, {O'Sullivan},
  {Ottens}, {Patapis}, {Penanen}, {Pietraszkiewicz}, {Ray}, {Regan},
  {Roteliuk}, {Royer}, {Samara-Ratna}, {Samuelson}, {Sargent}, {Scheithauer},
  {Schneider}, {Schreiber}, {Shaughnessy}, {Sheehan}, {Shivaei}, {Sloan},
  {Tamas}, {Teague}, {Temim}, {Tikkanen}, {Tustain}, {van Dishoeck},
  {Vandenbussche}, {Weilert}, {Whitehouse}, \& {Wolff}}]{wright23}
{Wright}, G.~S., {Rieke}, G.~H., {Glasse}, A., {et~al.} 2023, \pasp, 135,
  048003

\bibitem[{{Yan} {et~al.}(2010){Yan}, {Tacconi}, {Fiolet}, {Sajina}, {Omont},
  {Lutz}, {Zamojski}, {Neri}, {Cox}, \& {Dasyra}}]{yan10}
{Yan}, L., {Tacconi}, L.~J., {Fiolet}, N., {et~al.} 2010, \apj, 714, 100

\bibitem[{{Zhang} \& {Ho}(2023)}]{zhang23}
{Zhang}, L. \& {Ho}, L.~C. 2023, \apj, 943, 1

\bibitem[{{Zhang} {et~al.}(2022){Zhang}, {Ho}, \& {Li}}]{zhang22}
{Zhang}, L., {Ho}, L.~C., \& {Li}, A. 2022, \apj, 939, 22

\end{thebibliography}
\end{document}